# Trends in scientific research in Online Information Review. Part 2. Mapping the scientific knowledge through bibliometric and social network analyses

Juan Carlos Valderrama-Zurián[1]; Carolina Navarro-Molina[2]; Remedios Aguilar-Moya[2]; David Melero Fuentes[1]; Rafael Aleixandre-Benavent[14*]

1. Instituto de Documentación y Tecnologías de la Información, Universidad Católica de Valencia "San Vicente Mártir", Spain.
2. Unidad de Información e Investigación Social y Sanitaria (CSIC-UV), Spain.
3. Departamento de Ciencias de la Educación, Universidad Católica de Valencia "San Vicente Mártir", Spain.
4. Instituto de Gestión de la Información y del Conocimiento-Ingenio (CSIC-UPV), Spain.

**\*Correspondence:**
rafael.aleixandre@uv.es
Plaza Cisneros, 4
46003-Valencia, Spain

**Keywords:** Online Information Review journal; Bibliometric analysis; Social Network Analyses; Co-words analysis.

## Abstract

**Objective.** The purpose of this work is to analyse the knowledge structure and trends in scientific research in the *Online Information Reviews* journal by bibliometric analysis of key words and social network analysis of co-words.

**Methods.** Key words included in a set of 758 papers included in the Web of Science database from 2000 to 2014 were analysed. We conducted a subject analysis considering the key words assigned to papers. A social network analysis was also conducted to identify the number of co-occurrences between key words (co-words). The Pajek software was used to create and graphically visualize the networks.

**Results.** Internet is the most frequent key word (n=219) and the most central in the network of co-words, strongly associated with Information retrieval, search engines, the World Wide Web, libraries and users

**Conclusions**. Information science, as represented by *Online Information Review* in the present study, is an evolving discipline that draws on literature from a relatively wide range of subjects. Although *Online Information Review* appears to have well-defined and established research topics, the journal also changes rapidly to embrace new lines of research.

## 1. Background

In a previous paper, we conducted a bibliometric analysis of the journal *Online Information Review*, considering the scientific production, collaboration and citation patterns of authors, institutions and countries publishing in that journal. In the current work, we analyse the knowledge structure of scientific research of the journal, integrating the analyses of some characteristics of the bibliographic records as registered in the Web of Science database that are related to their thematic and conceptual content: authors' keywords, Keyword Plus and the networks of co-words. We used social network analysis to identify and graphically represent the existing networks of key words and co-words in the published papers.

There are several methods that may reveal the structure of a field and map the trends of scientific knowledge. Keywords Plus and Author Keywords are commonly selected as units of analysis, despite the limited research evidence demonstrating the effectiveness of Keywords Plus (Yang *et al*., 2012; Zhang *et al*., 2016). Author Keywords comprises a list of terms that authors consider best represent the content of their work whereas Keywords Plus is terms or phrases that appear frequently in the titles of an article's references and not necessarily in the title of the article itself or as Author Keywords. Keywords Plus is generated by an automatic computer algorithm (Garfield, 1990;

Garfield and Sher, 1993). According to Garfield (1990), Keywords Plus terms are able to capture an article's content with great depth and variety.

Developing co-words maps for the study of semantic relations in scientific literature was proposed and studied by Callon *et al*. (1983 and 1986) and Leydesdorff (1989 and 1997). Co-word analysis is a technique that is effective in mapping the strength of the association between information items in textual data. This technique directly addresses sets of keywords shared by publications, directly mapping the topics of the scientific literature from interactions of keywords (Hui and Fong, 2004; Wang *et al*., 2012). Co-word analysis is based on the assumption that key words compose an adequate description of the content of a paper. Thus, key words can be used to represent the content of a research field structure.

Our purpose, then, is to map scientific knowledge and identify research trends in Online Information Review by bibliometric and social network analyses. Identifying the knowledge structure of scientific research in this area can help neophytes become more familiar with this field, the specific thematic aspects addressed in the investigations, and their relations and interactions. An understanding of the knowledge structure may also provide appropriate understanding to progress in a new and competitively advantageous research direction.

## 2. Methods

### 2.1. Search strategy

This study follows the methodology of Aleixandre-Benavent *et al*. (2016). In our previous paper, we analysed scientific production, effects, and collaboration patterns in Online Information Review and identified research groups in a set of 758 papers included in the Web of Science database from 2000 to 2014. Of the 758 papers, 696 had key words and 428 had Keywords Plus. Only 60 papers did not have either type. We selected both types of key words.

## 2.2. Standarization of key words

To standardize key words, the core idea of simplicity was followed. Singulars and plurals (such as "catalogue" and "catalogues"), hyphenated words (such as "consumer-trust" and "consumer trust"), and spelling variants (such as "behavior" and "behaviour") were standardized. Some key words referring to the same idea were grouped. For example, "web site", "website", "site" and "web portal" were grouped under "web sites"; "citation counts" and "citation impact" were grouped under "citation analysis". Words with a similar meaning such as 'Hirsch index" and "h Index" or "journal impact factor" and "impact factor" were standardized as one term: in this case, 'h index" and "impact factor", respectively, were retained. After standardization, 1,612 different key words remained.

## 2.3. Bibliometric indicators

As previously described, we conducted a subject analysis considering the key words assigned to papers. A social network analysis was also conducted to identify the number of co-occurrences between key words (co-words). Co-occurrences refer to all combinations of key word pairs in each paper that are repeated in the set of papers revised. Co-word analysis has been effective in mapping the strength of the associations between key words in textual data. Social network analysis applied to co-word analysis allows us to draw network graphs that illustrate the strongest associations between descriptors (Lanza and Svendsen, 2007). For the analysis of co-words, we adopted the assumptions presented by Law and Whittaker (1992): a) authors of scientific papers choose their technical terms carefully; b) different terms are used in the same paper because the author recognizes some association between the terms; c) if different authors recognize the same relation, that relation may be assumed to have some significance within the area of science concerned.

The software Pajek (Batagelj and Mrvar, 2001) was used to create and graphically visualize the networks. The size of the spheres is proportional to the number of occurrences of each key word in the set of revised papers. Numbers in brackets indicate the number of papers that include the key word (figures 1 to 6) and the number of papers published by an institution (Figure 5) or country (Figure 6). The thickness of the lines connecting two spheres is proportional to the number of times these two key words appear simultaneously in the set of revised papers (figures 1 to 4) or the number of times the key word appears in the set of papers published by an institution (Figure 5) or country (Figure 6).

## 3. Results

### 3.1. Most frequent key words

Table 1 includes the distribution of papers according to the assigned key words in three five-year periods. Internet is the most frequently assigned key word (n=219), followed by World Wide Web (n=155), Information Retrieval (n=134) and Search Engines (n=106). Other key words that highlight with more than 50 published papers are Information (n=80) and Databases (n=57). Some key words increased during the analysed period (Figure 1), such as Information, Web Sites, Research, Trust, Search Engines and Electronic Commerce. Key words with an annual decreasing tendency of frequency (Figure 2) include Internet, Information Retrieval, Databases, Digital Libraries and Libraries.

*Figure 1. Annual evolution of most frequent key words with upward trend*

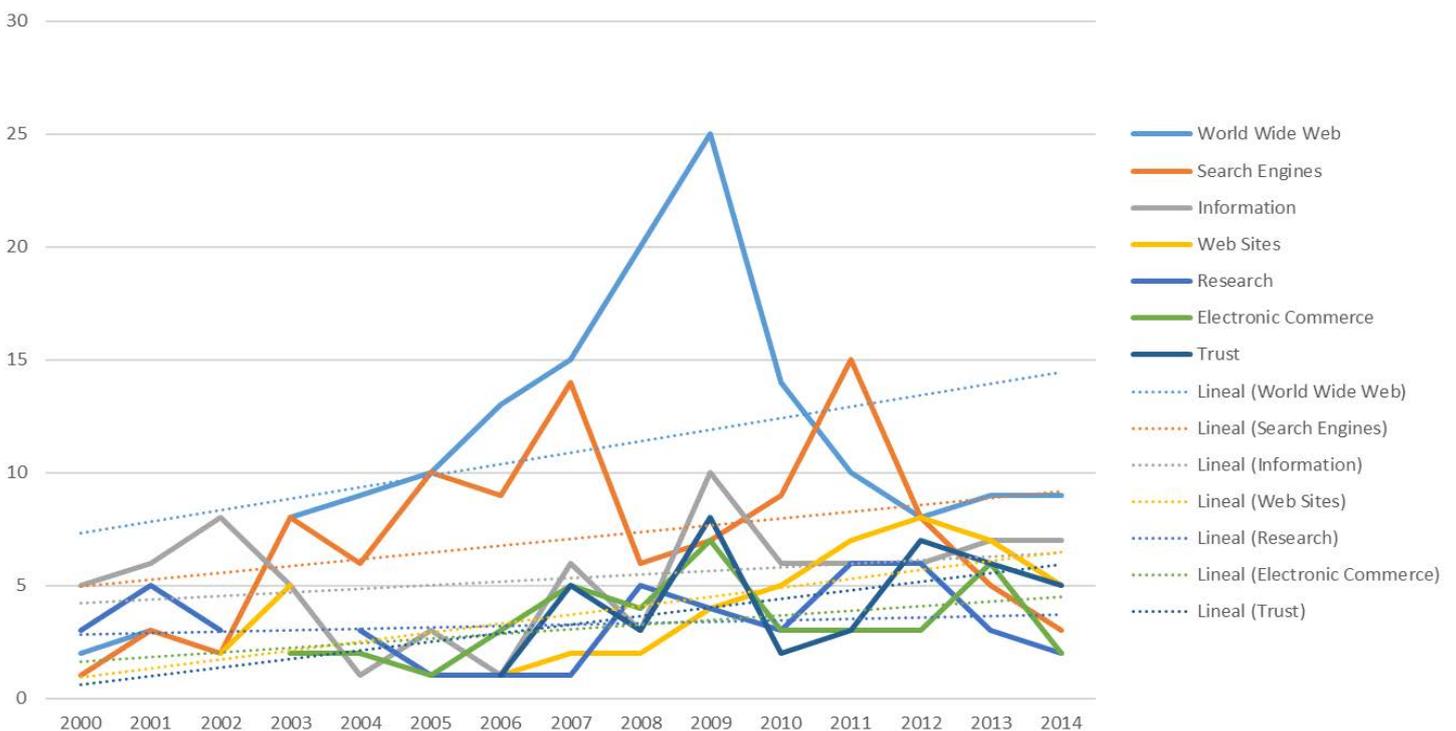

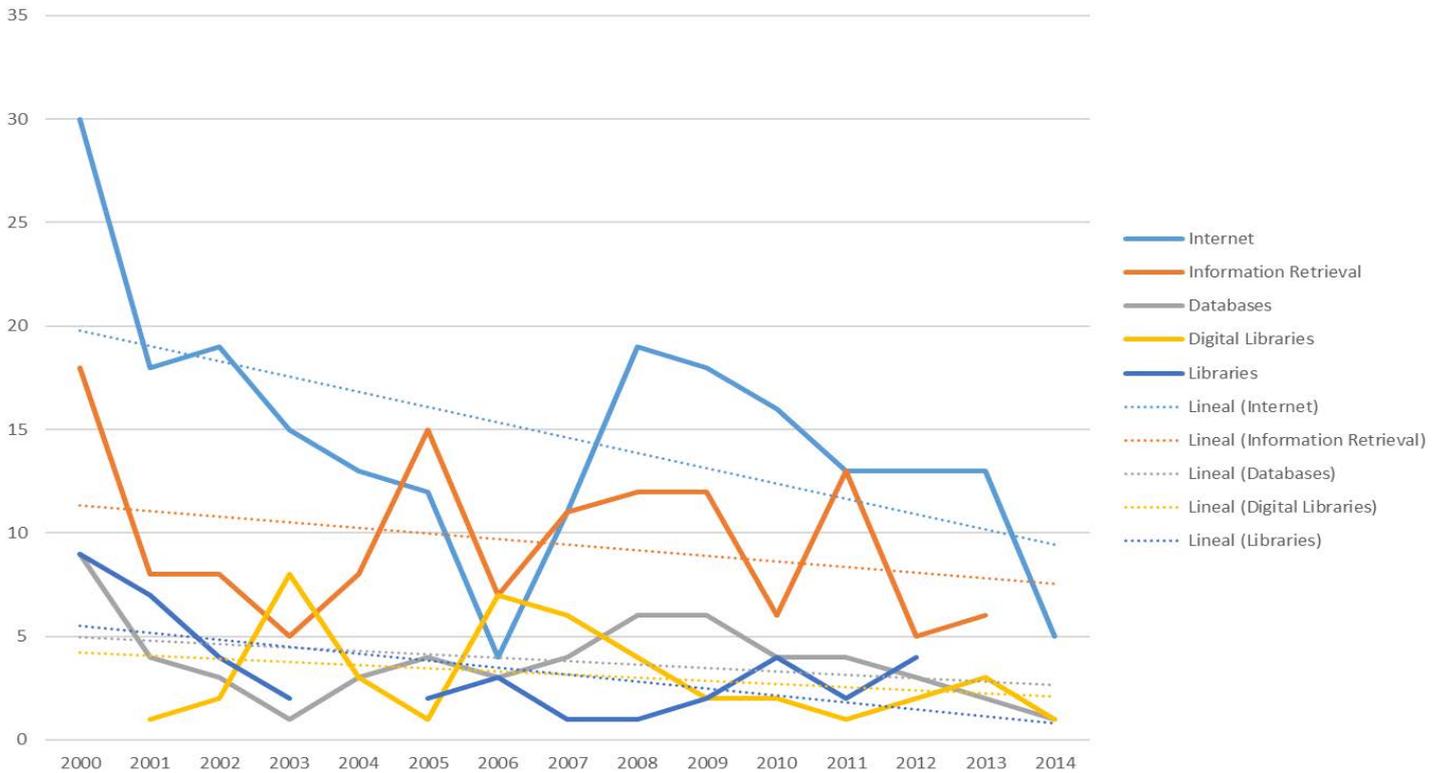

*Figure 2. Annual evolution of most frequent key words with downward trend*

## 3.2. Network of co-words

Figures 3 to 6 show the network of co-occurrence of key words (co-words) over the three considered periods and over the entire period. A threshold of more than 2 co-occurrences was established. During the first period, 2000-2004 (Figure 3), the network comprised 31 key words. Two key words occupied a more central position and a prominent intermediary position: Internet and information retrieval. Information retrieval occurred the most often (47 times); Internet is associated with 25 key words, followed by libraries (22 times) and databases (8 times). During the second period (2005-2009) (Figure 4), the network comprises 37 key words. During this period, World Wide Web is the key word with the most centrality because this phrase is connected to 16 other key words, followed by Internet, which is connected to other 14 key words, and information retrieval, with 9 key words. The closest associations were between information retrieval and search engines (n=19 times), World Wide Web and search engines (n=12 times), and World Wide Web and information retrieval (n=11 times). During the third period (2010-2014) (Figure 5), the network of co-words included 19

key words, and Internet and search engines were the key words with the most centrality and connections (n=11 times). Figure 6 shows the network of co-words during the entire period. To allow this graph to be correctly visualized, a threshold of more than 3 co-occurrences was established. The resultant network includes 56 key words. The network allows for perceiving the relations between the most central key words and the specific terms associated with those words. For example, Internet, the most central key word, is strongly associated with information retrieval (n=39 times), search engines (n=17), World Wide Web (n=14) and libraries (n=14) but also with other co-words such as users (n=10), information (n=9), databases (n=9), information services (n=9), social networks (n=9), information systems (n=9) and electronic commerce (n=9). Another central phrase, information retrieval, is strongly associated with Internet (n=39 times), search engines (n=35), databases (n=19) and World Wide Web (n=16) but also with another 20 phrases such as digital libraries (n=12), user studies (n=10), information searches (n=9) and research (n=7). World Wide Web is strongly associated with information retrieval (n=16), search engines (n=16), and Internet (n=14) but also with another 12 words such as cluster analysis (n=6), knowledge management (n=6) and semantics (n=6).

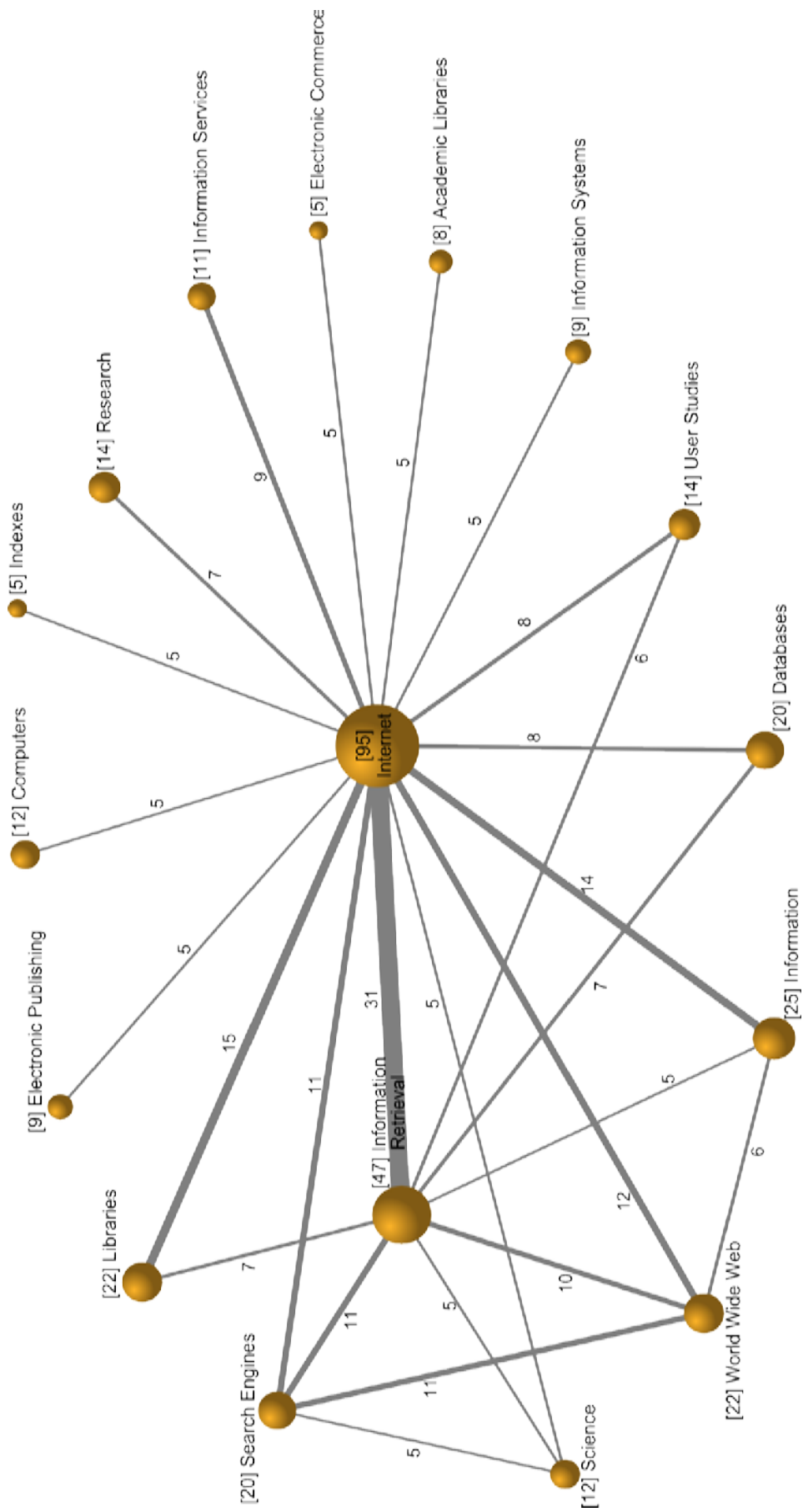

*Figure 3. Network of co-words in the 2000-2004 period*

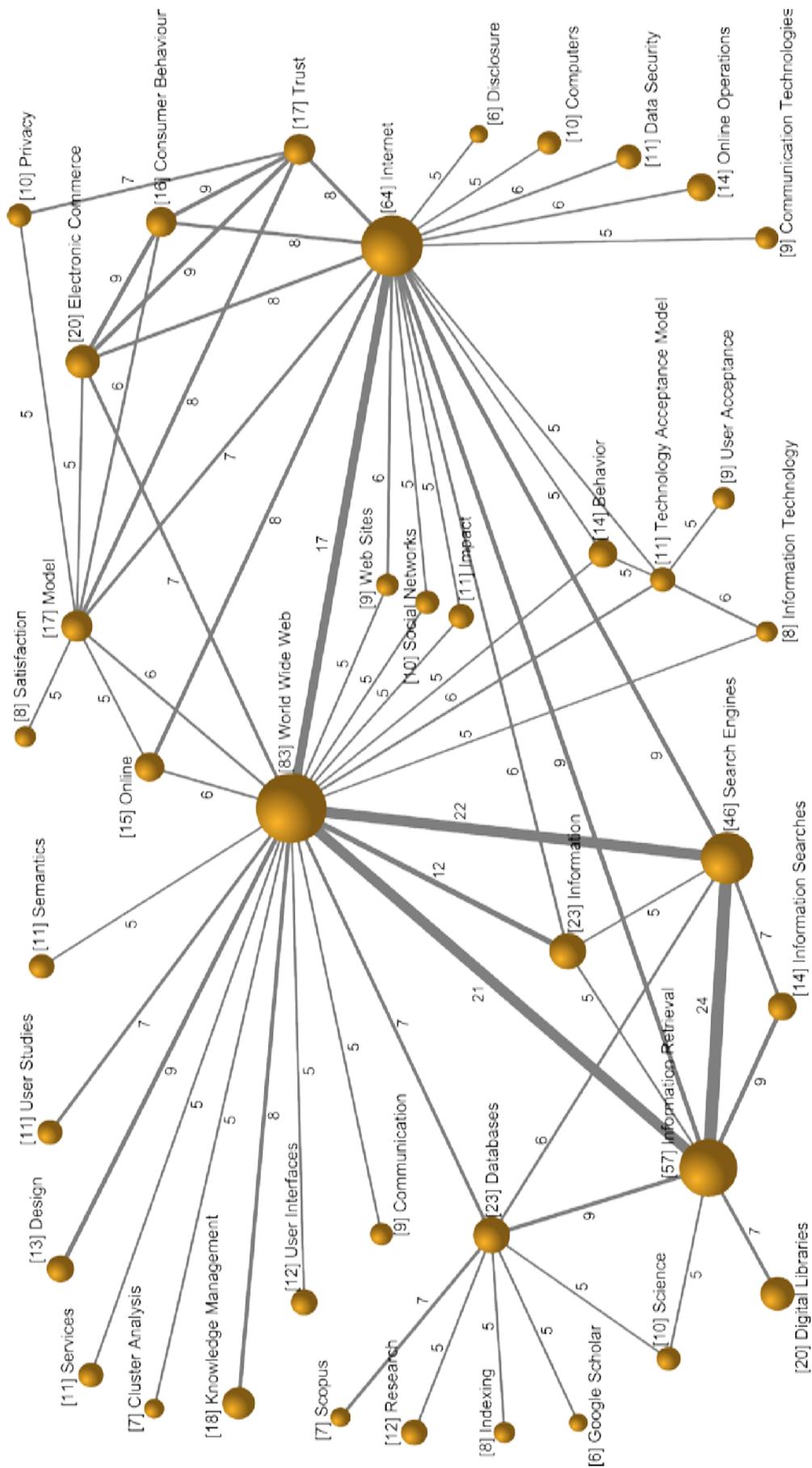

*Figure 4. Network of co-words in the 2005-2009 period*

*Figure 5. Network of co-words in the 2010-2014 period*

*Figure 6. Network of co-words along the complete analyzed period (2000-2014)*

## 3.3. Institutions publishing in *Online information Reviews* and associated key words

Figure 7 presents the key words most frequently associated with 12 institutions publishing more than 9 articles in *Online information Reviews*. A threshold of more than 2 occurrences was established. The most frequent key words included in papers published by the University of Hawaii Manoa (Hawaii, U.S.), the most productive institution with 77 published papers, were databases (n=27), *s*earch *e*ngines (n=20), information retrieval (n=18) and research (n=13). For the second most productive institution, Victoria University of Wellington (New Zealand, n=34), the most frequent key words were World Wide Web (n=11), Internet (n=6) and search engines (n=5). For Nanyang Technological University (Nanyang, Singapore), the third most productive institution (n=25), the most frequently associated key words were Internet (n=7) and information retrieval (n=7).

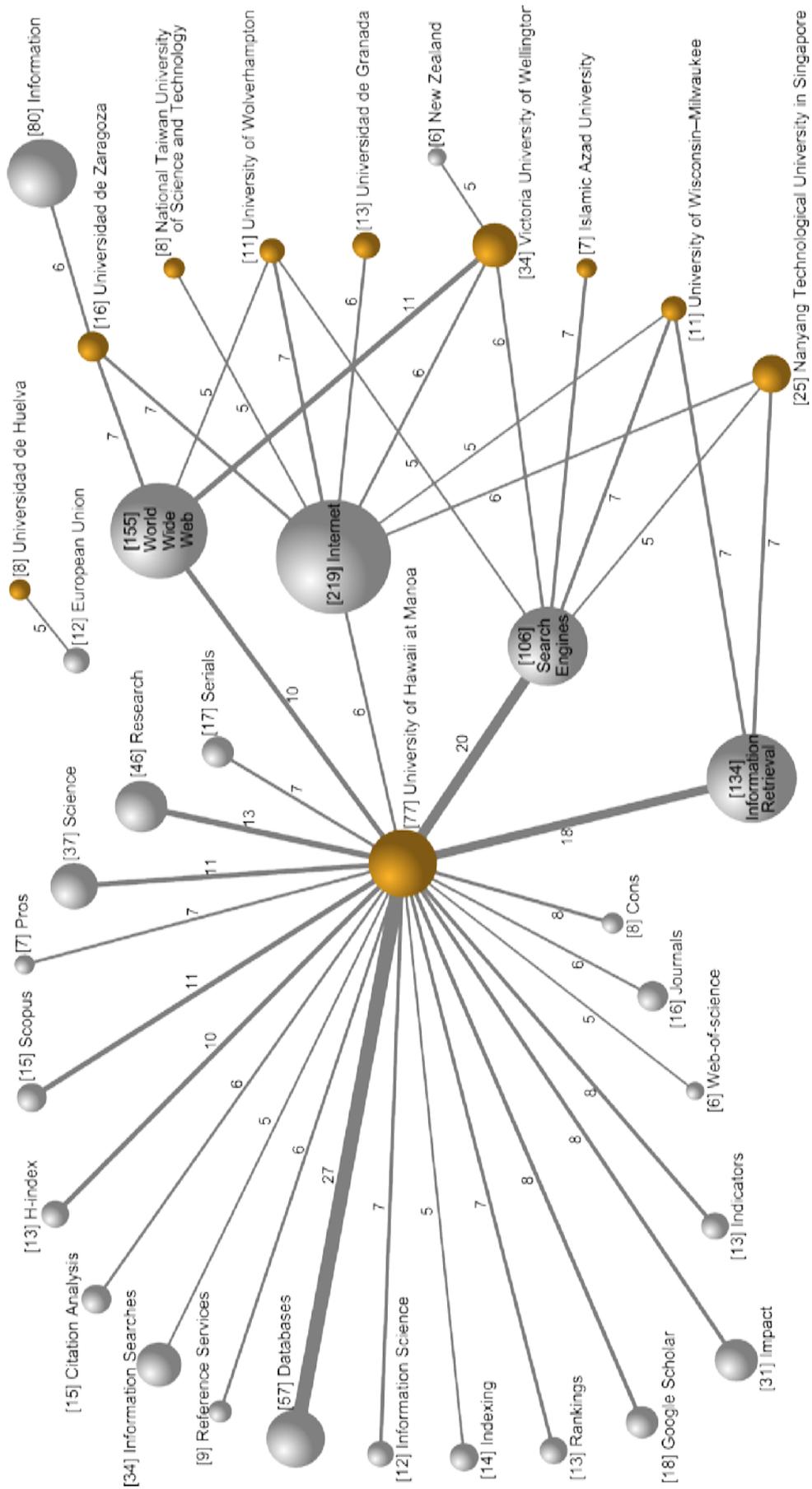

*Figure 7. Network of institutions and associated key words*

**3.4. Countries publishing in *Online information Reviews* and associated key words**

Figure 8 shows the key words most frequently associated with countries publishing more than 9 papers. A threshold of more than 4 occurrences was established. As seen in a previous work, of the 54 countries that contributed to the publication of papers, the United Sates ranked first with respect to scientific productivity (n=199). The most frequently used key words associated with this country were Internet (n=48), information retrieval (n=45), databases (n=33) and search engines (n=32). In second place for publishing the most papers, Taiwan (n=97) preferred Internet (n=26), World Wide Web (n=20), user acceptance (n=19) and technology acceptance model (n=18). In third place, Spain's (n=80) most used key words were Internet (n=31), World Wide Web (n=26), information (n=21), websites (n=14), electronic commerce (n=10) and disclosure (n=10).

*Figure 8. Network of countries and associated key words*

## 4. Discussion

This work presented the evolution of scientific knowledge on the research published in *Online Information Review* using the most-assigned key words for the papers and the SNA analysis of co-words. Key words are one of the best bibliometric indicators of the content of the papers, the primary concerns addressed in the articles and identifying growth in the subject field. To develop a more comprehensive knowledge of the topics addressed in the journal, we have analysed both Author Keywords and Keywords Plus. In this paper, we use the term keywords to refer to both types of words. The diffusion of knowledge and information regarding topics published in this journal may contribute to promoting a higher level of knowledge and cooperation within the information science community and creating a favourable environment for debate.

Several studies have been published that used bibliometric techniques and social network analyses of key words to map and assess knowledge in a particular field or topic of scientific research, including specific journals or a set of journals: consumer behaviour research (Muñoz-Leiva *et al*., 2012), stem cells (Ying and Wu, 2011), field associations terms (Rokaya *et al*., 2008), semantic mapping of words (Leydesdorff and Welbers, 2011), human intelligence network (Wang *et al*., 2012), ethics and dementia research (Baldwin *et al*., 2003), severe acute respiratory syndrome (Chiu *et al*., 2004), environmental science (Ho, 2007), adverse drug reactions (Clarke *et al*., 2007), tsunamis (Chiu and Ho, 2007), climate change (Li *et al*., 2011), and wine consumption and health (Aleixandre *et al*., 2013), among others. To compare and best observe tendencies by social network analyses of co-words, we have divided the 15 analysed years into three five-year periods.

As expected, both "classical" key words related to information systems and words related to the Internet environment predominate in the first five-year period analysed (2000 to 2004). During those years, two key words prevailed: Internet and information retrieval, both of which relate primarily to libraries, information services, search engines, user studies and databases. In the second period (2005-2009), the network of co-words was expanded, some words occupying a central position from the first period, such as Internet and information retrieval, World Wide Web and search engines. World Wide Web is clearly associated with search engines, information retrieval and Internet. The primary topics addressed by World Wide Web are associated with information, design and knowledge management whereas Internet focuses on electronic commerce, consumer behaviour, trust and model. Other prominent topics are related to data security, online operations and communication technologies. Some terms intermediate with others, such as database, which intermediates with Scopus, Research, Indexing and Google Scholar. During the third period (2010-2014), several developments are noticeable: the central position of the previous terms with the exception of information retrieval and the consolidation of user acceptance as a new central term. During this period, new terms appeared, such as online communities, virtual communities, and open systems, related to topics such as perception, intention, satisfaction, motivation and user acceptance. Other domains that have emerged and play a major role in stimulating research are investigations related to scientific publications based on bibliometric approaches to research evaluation, such as citation analysis, impact factor, h-index and rankings.

As seen, one of the leading novelties regarding research in the preceding decades is the significance of the World Wide Web, which draws numerous spheres of research together: computer applications, electronic media, electronic commerce, and information search and retrieval (González-Alcaide *et al*., 2008; Orduna-Malea *et al*., 2015). This consolidation is quite understandable because of the major influence of the development of the Internet as a technological tool, effecting a profound change in activities connected with library and information sciences (D'Elia *et al*., 2002; Tsay, 2004). One of the key words that is notable for not being a technical term in information science is trust, a term that is representative of the concern that the reliance on integrity, reliability and honesty have in the world of the Internet and the World Wide Web.

The presence of some key words, such as user acceptance, user studies, quality or consumer behaviour, in the top positions of frequency ranking reflects the change from preceding decades. Focus has shifted from organizations' internal aspects, their problems, and their role as intermediaries to valuing the role played by users as final consumers of information and the services offered (González-Alcaide *et al*., 2008; Jarvelin and Vakkari, 1990; Jarvelin and Vakkari, 1993; Kumpulainen, 1991; Xu and Yao, 2015). Increasingly, impressive growth has occurred in topics related to new technologies, databases and products, such as electronic commerce, social networks, Google Scholar, Scopus, open systems, virtual communities, and blogs. In summary, our results are consistent with previous studies, indicating that information science is an evolving discipline that draws on literature from a relatively wide range of subjects (Tsay, 2004; Tsay 2011). As in other works, our analyses of the literature published in *Online Information Review* revealed the interdisciplinary nature of information science envisioned by Borko (1968) and Saracevic (1999).

### 4.1. Limitations

Our study has some limitations that should be addressed. First, bibliometric and social network analyses were conducted on articles included in Web of Science. We determined that using Web of Science was preferable to using Scopus or Google Scholar based on several previous works (Vandermeulen *et al*., 2011; Meho and Yang, 2007). Second, conference proceedings were not included because the ideas presented in conferences are often republished in journals. However, it is possible that many of these papers were never published in refereed journals; therefore, such research was not analysed in our study. Third, because some records do not have assigned keywords, some topics may be underrepresented in our study. In addition, the quality of results from keyword analysis depends on a variety of factors, such as the quality of the keywords assigned by researchers and the accuracy with which these key words represent the content of the work. Some keywords may be too general, or authors may sometimes incorrectly emphasise topics that are not the most relevant in a research paper. General keywords may be useful in showing a rough overview of a scientific field but are less successful at representing detailed topics of a research area (Chen and Xiao, 2016).

### 4.2. Future research

Future research in this area could identify the evolution of traditional topics and the emergence of new areas of research. It also would be useful to compare topics published in *Online Information Review* with topics included in other library and information science journals to develop a broader view of the issues addressed in this field.

# 5. Conclusions

This work has provided helpful insights into the research published in *Online Information Review*. Many areas, including the most-used key words, the network of co-words and primary topics associated with institutions and countries, have been discussed. On the basis of the research findings, some conclusions and recommendations for the key network players are suggested. The scientometric and social network analysis of key words has great potential to expand valuable understanding of the advancement of the research in a specific journal as well as in cases of emerging fields such as new technologies and processes. This study clarifies the feasibility of co-word analysis as a viable method for extracting patterns and identifying trends in the research published in a given journal. Although estimating the amount and topics of research published in *Online Information Review* has not previously been reported, we believe that such a study is of primary importance to current and future discussions of the researcher's decisions regarding information science and library science. First, researchers possess useful information regarding the topics published in the journal. The journal's topics aligning with their own interest could guide researchers to select *Online Information Review* as a candidate journal for publication. Second, policy makers must have baseline information to estimate the status of specific journals to develop adequate policy measures. Finally, journal editors must know the state of current research to effectively position the journal in relation to their competitors (Vandermeulen *et al*., 2011).

In summary, our results reveal that information science, as represented by *Online Information Review* in the present study, is an evolving discipline that draws on literature from a relatively wide range of subjects. Although the journal appears to represent well-defined and established research topics, *Online Information Review* also changes rapidly to embrace new lines of research. Our findings should be of great interest to library and information scientists because such practitioners seek to understand the nature of research information science. In addition, it would be helpful for the journal editor to obtain the bibliometric portrait of the studied journal and recognize its interaction with other subject disciplines.